\documentclass[twocolumn, tighten]{aastex61}
\usepackage{amsmath}
\usepackage{listings}
\usepackage{pythonhighlight}
\shorttitle{Introduction to photon information efficiency (in bits per photon)}
\shortauthors{Michael Hippke}
\begin{document}
\title{Interstellar communication.\\V. Introduction to photon information efficiency (in bits per photon)}
\author{Michael Hippke}
\affiliation{Sonneberg Observatory, Sternwartestr. 32, 96515 Sonneberg, Germany}
\email{hippke@ifda.eu}

\begin{abstract}
How many bits of information can a single photon carry? Intuition says ``one'', but this is incorrect. With an alphabet based on the photon's time of arrival, energy, and polarization, several bits can be encoded. In this introduction to photon information efficiency, we explain how to calculate the maximum number of bits per photon depending on the number of encoding modes, noise, and losses.
\end{abstract}

\section{Introduction}
Interstellar communication is different from the one on Earth because the distances are so much larger, and the number of photons received so much lower. The distance to our nearest neighbor star ($10^{16}\,$m) is $10^{13}\times$ larger than the typical distance between a mobile phone and a base station ($10^3\,$m). Due to the inverse square law, a Watt-scale cm-sized beamed transmitter at GHz frequency can deliver $10^{20}$ photons per second to a meter-sized receiver over km distance, but only $10^{-13}$ photons per second over pc distance, about one per million years.

Consequently, photons are plentiful on Earth, but frequency channels are precious, because many mobile phone users hope to transfer large amounts of data simultaneously. Here, spectral efficiency (in bits per second per Hertz) is to be optimized. Conversely, even with large transmitters and receivers at high power, photons are precious over interstellar distance \citep{2017arXiv170603795H}, and here we aim to maximize the \textit{photon information efficiency} (in bits per photon) \citep{2014PhRvA..89d2309T}.

Intuitively, one would assume that a single photon can encode a digital ``0'' or ``1'' through its presence or absence, and that is about all the information a photon can hold. This is not correct, because one can associate a photon with a unique member $m$ of an alphabet (a \textit{mode}). In an alphabet with only two members (such as binary code), each member encodes one bit of information. For a larger alphabet, each photon carries more information.

In this paper, we present an introduction of how to calculate the number of bits per photon as a function of the alphabet size, noise, and losses. We will make use of easy examples and short computer code snippets.\\\\

\section{Encoding modes}
\label{modes}
The mapping of photons to an alphabet can be done by employing its dimensions. Photons can be characterized by their polarization (right or left circular polarization), orbital angular momentum, energy, and their time of arrival at a receiver.

\subsection{Exemplary encoding scheme}
For the time of arrival, we can use a simple signal modulation as an example, the pulse-position modulation. Each message bit is transmitted in one of $T=2^{k} ({k>0}, {k \in \mathbb{Z}}$) possible time slots (Table~\ref{tab1}). A single photon detected in one of $T$ slots then encodes log$_2 T$ bits of information. With $T=2^{3}=8$ slots, one photon encodes log$_2 (8)=3$ bits of information.

A similar scheme is possible with intensity modulation. Interchangeable descriptions for a photon's intensity (energy) are its wavelength and frequency, where the energy is $E=hc/ \lambda$, and $h$ is Planck's constant. The wavelength is defined as $\lambda=c/f$ with $c$ as the speed of light in vacuum. A simple encoding scheme based on different photon energies could again define $E=2^{i} ({i>0}, {i \in \mathbb{Z}}$) possible energy levels, and achieve log$_2 E$ bits of information per photon.

Finally, encoding modes can be combined. We can imagine a receiver which detects photons in 8 distinct time slots every second, and measures each photon's polarization, as well as its energy at 8 distinct levels. Neglecting polarization, this is possible with Earth 2017 technology using time-resolved spectroscopy, at a receiver efficiency of about 10--20\%. With 8 modes from the time of arrival, 8 from the energy of the photon, and two from its polarization, we have $m=8\times8\times2=128$ modes available per photon.

The inverse of this quantity, $M=m^{-1}$, is the \textit{number of photons per mode}. If we send one photon per second, we have $M=1/128\approx0.008$.\\

\begin{table}
\center
\caption{Pulse-position modulation}
\label{tab1}
\begin{tabular}{cccccccc}
$\square$ & $\square$ & $\square$ & $\square$ & $\square$ & $\boxtimes$ & $\square$ & $\square$ \\
1&2&3&4&5&6&7&8
\end{tabular}
\end{table}

\section{Losses}
Some fraction of photons is lost between transmitter and receiver. For example, atmospheric transmission depends on the wavelength and varying characteristics, such as the content of water vapor in the air. It varies between essentially zero transmission, e.g. for X-rays, and very high ($>99$\%) in some infrared bands.

For space-based communication, extinction causes a loss of optical photons from the scattering\footnote{For simplicity, we neglect other disturbances such as scatter broadening.} of radiation by dust \citep{1996Ap&SS.236..285R}. Extinction is small ($<1$\%) for short distances (pc) at optical wavelengths, but increases with distance to $>10$\% at 200\,pc \citep{1998A&A...340..543V}.

Finally, the receiver itself has limited efficiency $<100$\%. The surviving part of the flux is noted as the \textit{transmissivity} $\eta$ for further treatment in section~\ref{h}.

\section{Noise}
Major sources of noise photons are instrumental, the atmospheric sky background (when observing from a planet), zodiacal light, reflections from celestial bodies (e.g., the moon), starlight (galactic and extragalactic), and the cosmic microwave background. Each noise source has its own wavelength characteristics. In total, the darkest observatory sites on earth have an optical sky background of $22$\,mag\,arcsec$^{-2}$ \citep{2008ASPC..400..152S}, corresponding to an optical flux of a few photons arcsec$^{-2}$\,sec$^{-1}$.

When communicating with a probe on an interstellar mission, target starlight might get blended with the probe's communication beam. For a flight to Proxima Cen, the sky-projected distance is of order arcsec at a separation of 1\,au from the star\footnote{During most of the flight, the problem is much less severe because of the large proper motion of $\alpha$\,Cen \citep[3.7\,arcsec\,yr$^{-1}$,][]{2016A&A...594A.107K}.}, and might allow for the use of coronographs to suppress $10^{-9}$ \citep{2006ApJS..167...81G,2015RAA....15..453L} of the starlight ($4\times10^{6}$ photons sec$^{-1}$\,m$^{-2}$).

Finally, the receiver itself will introduce a small amount of thermal noise into the detector.

All relevant noise sources can be added up for a specific communication at some wavelength (range). If the noise flux is uniform over the communication band, we can divide the noise flux by the number of modes (see section~\ref{modes}) for further calculations in section~\ref{h}. The \textit{average number of noise photons per mode} is noted as $N_{\rm M}$. If the flux is not monochromatic and uniform, calculations can be done step-wise.

\section{Calculating bits per photon}
\label{h}
In theory, free-space optical communication has unbounded photon information efficiency \citep{2011arXiv1104.2643D}. In practice, limitations arise from the limited number of modes $M$, a certain number of noise photons per mode $N_{\rm M}$, and the transmissivity $\eta$. Holevo's bound \citep{holevo1973bounds} establishes the upper limit to the amount of information which can be transmitted with a quantum state.

\begin{figure}
\includegraphics[width=\linewidth]{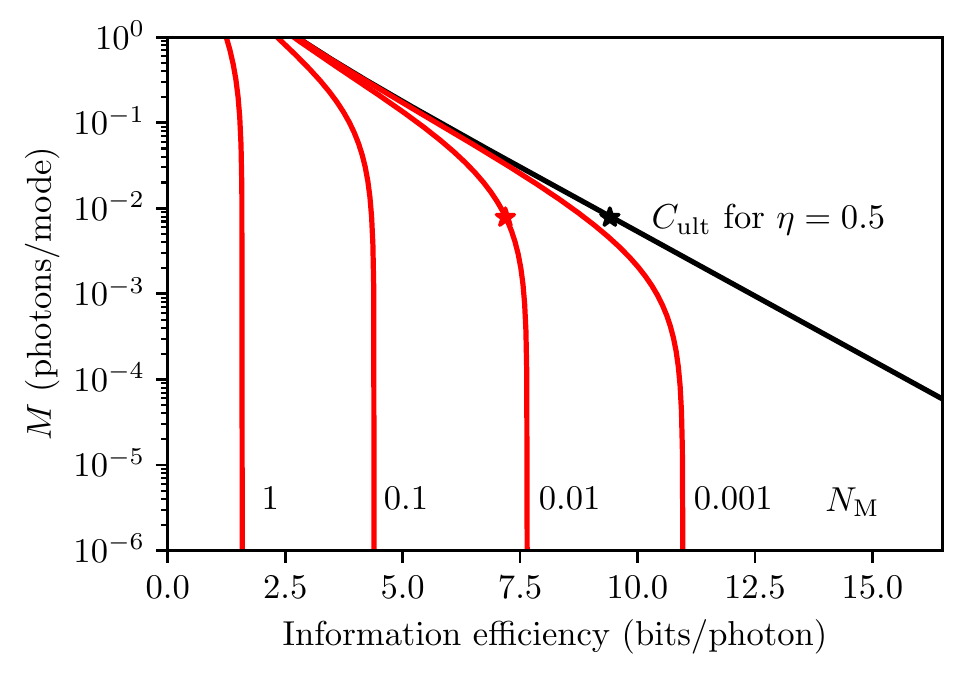}
\caption{\label{figure_h1}Information efficiency as a function of the number of photons per mode. More modes can encode more bits per photon, however the ultimate bound (black) is logarithmic. When accounting for thermal noise per mode $N_{\rm M}$ (fractions in the plot), the limits are even tighter (red lines). The examples from section~\ref{h} are shown with black (noiseless) and red (noisy) symbols.}
\end{figure}

\subsection{Holevo's bound without noise}
As shown by \citet{2004PhRvL..92b7902G}, the ultimate noiseless quantum limit can be expressed as:

\begin{equation}
C_{\rm ult}=g(\eta M) \,\,\,{\rm (bits\,\,per\,\,photon)}
\end{equation}

where

\begin{equation}
\label{gx}
g(x)=(1+x) \log_2 (1+x)-x \log_2 x
\end{equation}

so that $g(x)$ is a function of $\eta \times M$ \citep{2014PhRvA..89d2309T}. The unwieldy expression can be written as a \texttt{Python} code function:

\begin{python}
def holevo_perfect(M, eta):
    return ((1 + M * eta) * log2(1 + M * eta) -
        M * eta * log2(M * eta)) / (M * eta)
\end{python}

For $M=0.008$, $\eta=0.5$, the function evaluates to

\begin{python}
print(holevo_perfect(M=0.008, eta=0.5))
--> 9.41 (bits per photon)
\end{python}

\subsection{Holevo's bound with noise}
With noise, it was conjectured \citep{2004PhRvA..70c2315G} and recently proven \citep{2014NaPho...8..796G} that the limit is:

\begin{equation}
\label{thermal_holevo}
C_{\rm th}=g(\eta M + (1-\eta) N_{\rm M}) - g((1-\eta)N_{\rm M})
\end{equation}

Again, we can write the expression in \texttt{Python} code:
\begin{python}
def holevo_th(M, N_th, eta):
    def gx(x, eta, M):
        return (((1 + x)) * log2(1 + x) -
            x * log2(x)) / (M * eta)
    a = M * eta + (1 - eta) * N_th
    b = (1 - eta) * N_th
    return gx(a, eta, M) - gx(b, eta, M)
\end{python}

For $M=0.008$, $N_{\rm M}=0.01$, $\eta=0.5$, the function evaluates to

\begin{python}
print(holevo_th(M=0.008, N_th=0.01, eta=0.5))
--> 7.19 (bits per photon)
\end{python}

which is below the ultimate limit due to noise.

\section{Discussion and outlook}

\subsection{Photon Orbital Angular Momentum}
The ``Photon Orbital Angular Momentum'' (POAM) was recently found to be able to encode information on individual photons \citep{2002PhRvL..88y7901L,2003ApJ...597.1266H}. In practice, photons have been made with OAMs as high as $\ell = 200 \hbar$. Unfortunately, higher-order OAMs spatially widen the beam by $\sqrt{m}$ with m as the number of modes \citep{2003OptCo.223..117W}. The data rate is an (inverse) quadratic function of the beam width \citep{2017arXiv170603795H}, but only a logarithmic function of the number of modes (section~\ref{h}). Therefore, it is never preferable to use $\ell >1$ in photon limited free-space communications. We neglect such modulations here, although they might be useful in short-distance communications where beam widths may be large.

The same issue applies to an observer, whose sky-projected resolution for increasing $\ell$ becomes ever larger at a given aperture, also increasing noise. High-$\ell$ observations will require implausibly large apertures.

\subsection{Realistic encoding schemes}
It is an open question if the Holevo bound can fully, or only approximately be achieved in practice \citep{2012arXiv1202.0518W,2012arXiv1202.0533G}. It appears that the limit can be approached with more and more complicated encoding schemes \citep{2011arXiv1104.2643D}. If a reasonably simple scheme were found to be optimal, we would employ it for our own communications, and have reason to believe it would be widely used by putative other civilizations.

As of now, practical encoding schemes (with photon counting) come within a factor of a few of the bound. Coherent communication systems with carrier phase information at the receiver (not photon counting) apply matched filters for information retrieval. In this classical scheme, the well-known \citep{S1949} upper bound applies with a maximum achievable photon efficiency of 1.44 (2.88) bits/photon for heterodyne (homodyne) receivers \citep{majumdar2014advanced}.

\subsection{Signal to noise ratio}
Figure~\ref{figure_h1} shows a prominent knee in each noisy curve, representing a cut-off where the information efficiency (in bits per photon) can not be increased further, despite using more and more encoding modes (fewer photons per mode). Intuitively, this represents the point where the signal photons get overwhelmed by noise photons, $S \ll N$. As a rule of thumb, the logarithmic information efficiency improvement can be leveraged for $S>N$.

\subsection{Logarithmic advantage of the number of modes}
As can be seen from Figure~\ref{figure_h1} and Eq.~\ref{gx}, increasing the number of modes only logarithmically increases the information efficiency (in bits per photon). Even for very low noise levels, very few signal photons, and many modes, the information efficiency will not exceed more than a few dozen bits per photon. As an extreme example, consider the noiseless case with one transmitted photon, $M=10^{20}$ modes at $\eta=0.99$, which yields 66 bits per photon. Even very advanced technology will be limited to about 10 bits per photon in interstellar communications by noise and inevitable losses.

\subsection{Theoretical and practical limits}
One limit to the information efficiency comes from thermal noise in the universe, which is a function of cosmic time. Another limit is on the number of modes per unit time which arises from Heisenberg's uncertainty principle and will be examined in detail in a subsequent paper of this series.

\acknowledgments
\texttt{Acknowledgments}
The author is thankful to Jason T. Wright and Dainis Dravins for discussions on photon orbital angular momentum.

\pagebreak

\bibliographystyle{yahapj}

\begin{thebibliography}{}
\providecommand\natexlab[1]{#1}
\providecommand\JournalTitle[1]{#1}

\bibitem[{{Dolinar} {et~al.}(2011){Dolinar}, {Birnbaum}, {Erkmen}, \&
  {Moision}}]{2011arXiv1104.2643D}
{Dolinar}, S., {Birnbaum}, K.~M., {Erkmen}, B.~I., \& {Moision}, B. 2011,
  \JournalTitle{ArXiv e-prints},
  \href{http://arxiv.org/abs/1104.2643}{{\sffamily arXiv:1104.2643 [quant-ph]}}

\bibitem[{{Giovannetti} {et~al.}(2014){Giovannetti},
  {Garc{\'{\i}}a-Patr{\'o}n}, {Cerf}, \& {Holevo}}]{2014NaPho...8..796G}
{Giovannetti}, V., {Garc{\'{\i}}a-Patr{\'o}n}, R., {Cerf}, N.~J., \& {Holevo},
  A.~S. 2014,
  \href{http://dx.doi.org/10.1038/nphoton.2014.216}{\JournalTitle{Nature
  Photonics}, 8, 796}

\bibitem[{{Giovannetti} {et~al.}(2004{\natexlab{a}}){Giovannetti}, {Guha},
  {Lloyd}, {Maccone}, \& {Shapiro}}]{2004PhRvA..70c2315G}
{Giovannetti}, V., {Guha}, S., {Lloyd}, S., {Maccone}, L., \& {Shapiro}, J.~H.
  2004{\natexlab{a}},
  \href{http://dx.doi.org/10.1103/PhysRevA.70.032315}{\JournalTitle{\pra}, 70,
  032315}

\bibitem[{{Giovannetti} {et~al.}(2004{\natexlab{b}}){Giovannetti}, {Guha},
  {Lloyd}, {Maccone}, {Shapiro}, \& {Yuen}}]{2004PhRvL..92b7902G}
{Giovannetti}, V., {Guha}, S., {Lloyd}, S., {et~al.} 2004{\natexlab{b}},
  \href{http://dx.doi.org/10.1103/PhysRevLett.92.027902}{\JournalTitle{Physical
  Review Letters}, 92, 027902}

\bibitem[{{Guha} \& {Wilde}(2012)}]{2012arXiv1202.0533G}
{Guha}, S., \& {Wilde}, M.~M. 2012, \JournalTitle{ArXiv e-prints},
  \href{http://arxiv.org/abs/1202.0533}{{\sffamily arXiv:1202.0533 [cs.IT]}}

\bibitem[{{Guyon} {et~al.}(2006){Guyon}, {Pluzhnik}, {Kuchner}, {Collins}, \&
  {Ridgway}}]{2006ApJS..167...81G}
{Guyon}, O., {Pluzhnik}, E.~A., {Kuchner}, M.~J., {Collins}, B., \& {Ridgway},
  S.~T. 2006, \href{http://dx.doi.org/10.1086/507630}{\JournalTitle{\apjs},
  167, 81}

\bibitem[{{Harwit}(2003)}]{2003ApJ...597.1266H}
{Harwit}, M. 2003, \href{http://dx.doi.org/10.1086/378623}{\JournalTitle{\apj},
  597, 1266}

\bibitem[{{Hippke}(2017)}]{2017arXiv170603795H}
{Hippke}, M. 2017, \JournalTitle{ArXiv e-prints},
  \href{http://arxiv.org/abs/1706.03795}{{\sffamily arXiv:1706.03795
  [astro-ph.IM]}}

\bibitem[{Holevo(1973)}]{holevo1973bounds}
Holevo, A.~S. 1973, \JournalTitle{Problemy Peredachi Informatsii}, 9, 3

\bibitem[{{Kervella} {et~al.}(2016){Kervella}, {Mignard}, {M{\'e}rand}, \&
  {Th{\'e}venin}}]{2016A&A...594A.107K}
{Kervella}, P., {Mignard}, F., {M{\'e}rand}, A., \& {Th{\'e}venin}, F. 2016,
  \href{http://dx.doi.org/10.1051/0004-6361/201629201}{\JournalTitle{\aap},
  594, A107}

\bibitem[{{Leach} {et~al.}(2002){Leach}, {Padgett}, {Barnett}, {Franke-Arnold},
  \& {Courtial}}]{2002PhRvL..88y7901L}
{Leach}, J., {Padgett}, M.~J., {Barnett}, S.~M., {Franke-Arnold}, S., \&
  {Courtial}, J. 2002,
  \href{http://dx.doi.org/10.1103/PhysRevLett.88.257901}{\JournalTitle{Physical
  Review Letters}, 88, 257901}

\bibitem[{{Liu} {et~al.}(2015){Liu}, {Ren}, {Dou}, {Zhu}, {Zhang}, {Zhao},
  {Wu}, \& {Chen}}]{2015RAA....15..453L}
{Liu}, C.-C., {Ren}, D.-Q., {Dou}, J.-P., {et~al.} 2015,
  \href{http://dx.doi.org/10.1088/1674-4527/15/3/012}{\JournalTitle{Research in
  Astronomy and Astrophysics}, 15, 453}

\bibitem[{Majumdar(2014)}]{majumdar2014advanced}
Majumdar, A. 2014, Advanced Free Space Optics (FSO): A Systems Approach,
  Springer Series in Optical Sciences (Springer New York)

\bibitem[{{Ryter}(1996)}]{1996Ap&SS.236..285R}
{Ryter}, C.~E. 1996,
  \href{http://dx.doi.org/10.1007/BF00645150}{\JournalTitle{\apss}, 236, 285}

\bibitem[{Shannon \& Weaver(1949)}]{S1949}
Shannon, C.~E., \& Weaver, W. 1949, The Mathematical Theory of Communication
  (University of Illinois press), vii + 117

\bibitem[{{Smith} {et~al.}(2008){Smith}, {Warner}, {Orellana}, {Munizaga},
  {Sanhueza}, {Bogglio}, \& {Cartier}}]{2008ASPC..400..152S}
{Smith}, M.~G., {Warner}, M., {Orellana}, D., {et~al.} 2008, in Astronomical
  Society of the Pacific Conference Series, Vol. 400, Preparing for the 2009
  International Year of Astronomy: A Hands-On Symposium, ed. M.~G. {Gibbs},
  J.~{Barnes}, J.~G. {Manning}, \& B.~{Partridge}, 152

\bibitem[{{Takeoka} \& {Guha}(2014)}]{2014PhRvA..89d2309T}
{Takeoka}, M., \& {Guha}, S. 2014,
  \href{http://dx.doi.org/10.1103/PhysRevA.89.042309}{\JournalTitle{\pra}, 89,
  042309}

\bibitem[{{Vergely} {et~al.}(1998){Vergely}, {Ferrero}, {Egret}, \&
  {Koeppen}}]{1998A&A...340..543V}
{Vergely}, J.-L., {Ferrero}, R.~F., {Egret}, D., \& {Koeppen}, J. 1998,
  \JournalTitle{\aap}, 340, 543

\bibitem[{{Wei} {et~al.}(2003){Wei}, {Xue}, {Leach}, {Padgett}, {Barnett},
  {Franke-Arnold}, {Yao}, \& {Courtial}}]{2003OptCo.223..117W}
{Wei}, H., {Xue}, X., {Leach}, J., {et~al.} 2003,
  \href{http://dx.doi.org/10.1016/S0030-4018(03)01619-5}{\JournalTitle{Optics
  Communications}, 223, 117}

\bibitem[{{Wilde} {et~al.}(2012){Wilde}, {Guha}, {Tan}, \&
  {Lloyd}}]{2012arXiv1202.0518W}
{Wilde}, M.~M., {Guha}, S., {Tan}, S.-H., \& {Lloyd}, S. 2012,
  \JournalTitle{ArXiv e-prints},
  \href{http://arxiv.org/abs/1202.0518}{{\sffamily arXiv:1202.0518 [quant-ph]}}

\end{thebibliography}

\end{document}